\documentclass[final]{IEEEtran}

\usepackage{xcolor,soul,framed} 

\colorlet{shadecolor}{yellow}

\usepackage[cmex10]{amsmath}
\usepackage{array}
\usepackage{ amssymb }
\usepackage{mdwmath}
\usepackage[noadjust]{cite}
\usepackage{mdwtab}
\usepackage{amsmath}
\usepackage{eqparbox}
\usepackage{tabularx,booktabs}
\newcolumntype{C}{>{\centering\arraybackslash}X} 
\usepackage{graphicx}
\ifCLASSOPTIONcompsoc
    \usepackage[caption=false, font=normalsize, labelfont=sf, textfont=sf]{subfig}
\else
\usepackage[caption=false, font=footnotesize]{subfig}
\fi
\usepackage{multicol}
\usepackage{url}

\begin{document}
    \title{High-Performance All-Optical Modulator Based on Graphene-hBN Heterostructures}
  \author{Mohammed Alaloul and Jacob B. Khurgin

  \thanks{Mohammed Alaloul was with New York University Abu Dhabi, Abu Dhabi, UAE. He is now with the School of Engineering and Information Technology at the University of New South Wales, Canberra, ACT 2612, Australia (e-mail: m.alaloul@unsw.edu.au). Jacob B. Khurgin is with the Electrical and Computer Engineering Department, Johns Hopkins University, Baltimore, MD 21218, USA (e-mail: jakek@jhu.edu).}
 }


\maketitle

\begin{abstract}

Graphene has emerged as an ultrafast photonic material for on-chip all-optical modulation. However, its atomic thickness limits its interaction with guided optical modes, which results in a high switching energy per bit or low modulation efficiencies. Nonetheless, it is possible to enhance the interaction of guided light with graphene by nanophotonic means. Herein, we present a practical design of an all-optical modulator that is based on graphene and hexagonal boron nitride (hBN) heterostructures that are hybrid integrated into silicon slot waveguides. Using this device, a high extinction ratio (ER) of 7.3$\,$dB, an ultra-low insertion loss (IL) of $<$0.6$\,$dB, and energy-efficient switching ($<$0.33$\,$pJ/bit) are attainable for a 20$\,\text{\textmu}$m long modulator with double layer graphene. In addition, the device performs ultrafast switching with a recovery time of $<$600$\,$fs, and could potentially be employed as a high-performance all-optical modulator with an ultra-high bandwidth in the hundreds of GHz. Moreover, the modulation efficiency of the device is further enhanced by stacking additional layers of graphene-hBN heterostructures, while theoretically maintaining an ultrafast response. The proposed device exhibits highly promising performance metrics, which are expected to serve the needs of next-generation photonic computing systems.

\end{abstract}


%
\IEEEpeerreviewmaketitle


\section{Introduction}

\begin{table*}
\label{tab}
 \caption{Fixed parameters in the design of the on-chip all-optical modulator.}
\label{my-label}
\begin{tabularx}{\textwidth}{@{}l*{8}{C}c@{}}
\toprule
\textbf{Parameter} & $h$ & $d$ & $W$ & $W_0$ & $w_1$ & $w_2$ & $L_{\text{tap}}$ \\ 
\midrule
\textbf{Value} & 250$\,$nm & 80$\,$nm & 250$\,$nm & 400$\,$nm & 60$\,$nm & 30$\,$nm & 8$\,$\textmu m  \\ 
\bottomrule
\end{tabularx}
\end{table*}

\IEEEPARstart{A}{ll-Optical} modulators are devices that control the flow of light in a photonic link by applying a time-varying optical signal \cite{chen2020all}. Unlike electro-optic modulators, these devices modulate light without performing optical-electrical-optical conversions that lead to additional radio-frequency time delays in the signal processing \cite{rutckaia2020ultrafast}. Moreover, they are characterized by ultrafast switching times ($<$$1\,$ps) \cite{chai2017ultrafast}, and are hence employed for applications in photonic computing, optical logic and optical information processing \cite{sasikala2018all}. However, these modulators consume several picojoules of energy per bit \cite{neira2014ultrafast, huang2018sigec,li2007ultrafast}, whereas next-generation photonic devices for telecom and datacom networks require energy-efficient devices with a switching energy that is $<$$1 \,$pJ/bit \cite{giambra2021wafer}. Energy-efficient modulators with a sub-pJ switching energy have been demonstrated in refs. \cite{nozaki2010sub, husko2009ultrafast, takiguchi2020all}, but were limited by relatively high switching times (\textgreater 1$\,$ps). Recently, an all-optical device that is characterized by ultrafast (260$\,$fs) and energy-efficient (35$\,$fJ) switching has been demonstrated in \cite{ono2020ultrafast}. This device is based on graphene-loaded deep-subwavelength plasmonic waveguides, where the use of plasmonic structures enhanced the interaction of the guided optical mode with graphene, leading to a significant reduction in the switching energy. Other plasmon-enhanced all-optical graphene modulators have been reported in refs. \cite{sun2019all, alaloul2021low}. Nonetheless, incorporating plasmonic metals into these devices introduces an excessive insertion loss ($IL$), mainly due to the inherent ohmic losses that are induced by these metals \cite{maier2007plasmonics}. Consequently, a greater input optical power would be required to achieve a decent signal-to-noise ratio (SNR) at the receiver side, which might hinder the adoption of these devices in next-generation optical interconnects. Graphene is a promising candidate material for next-generation all-optical modulators because of its ultrafast electron heating and cooling dynamics \cite{tielrooij2015generation, alaloul2021plasmon, chen2019highly}, facile integration into silicon photonics \cite{gan2013chip,youngblood2014multifunctional,phare2015graphene}, and compatibility with complementary metal oxide semiconductor (CMOS) processes \cite{pospischil2013cmos, akinwande2019graphene,jiang2018cmos}. Nevertheless, building energy-efficient all-optical graphene modulators is challenging because of graphene's atomic thickness, which limits its interaction with guided optical modes. Therefore, alternative nanophotonic structures are needed to enhance the interaction of waveguide modes with graphene, while maintaining a low insertion loss in the photonic link. In this work, we propose an all-optical modulator that is based on heterostructures of graphene and hexagonal boron nitride (hBN) that are hybrid integrated into silicon slot waveguides. The use of graphene-hBN heterostructures accelerates the electron cooling dynamics in graphene \cite{tielrooij2018out, principi2017super, huang2020ultra, golla2017ultrafast}, and facilitates the stacking of multiple monolayers of high-quality graphene \cite{gannett2011boron,dean2010boron, zomer2011transfer}, while the use of slot waveguides boosts the interaction of the guided optical mode with graphene \cite{wang2016high, cheng2015graphene, wang2016plane}. The proposed device performs energy-efficient switching ($<$$0.33\,$pJ/bit) with a high extinction ratio (ER) of 7.3$\,$dB at an ultra-low IL of $<$$0.6\,$dB for a 20$\,\text{\textmu}$m long modulator with double layer graphene. Furthermore, its switching performance is ultrafast with a theoretical recovery time of $<$$600\,$fs, respectively, and could potentially be employed as a high-performance all-optical modulator with an ultra-high bandwidth in the hundreds of GHz. \\
\indent In the next section, the device structure and its constituent materials are presented. Following that, the all-optical modulation mechanism is introduced and explained, and the device efficiency is characterized by calculating its switching energy and $ER$. Then, the same quantities are studied at other wavelengths to characterize the broadband response of the modulator. Subsequently, the device performance is investigated by calculating the rise and fall times of the modulator. Finally, the report is concluded by pointing out the major findings of this study.

\section{Device Structure}

\begin{figure} 
    \centering
  \subfloat[\label{1a}]{%
       \includegraphics[width=0.7\linewidth]{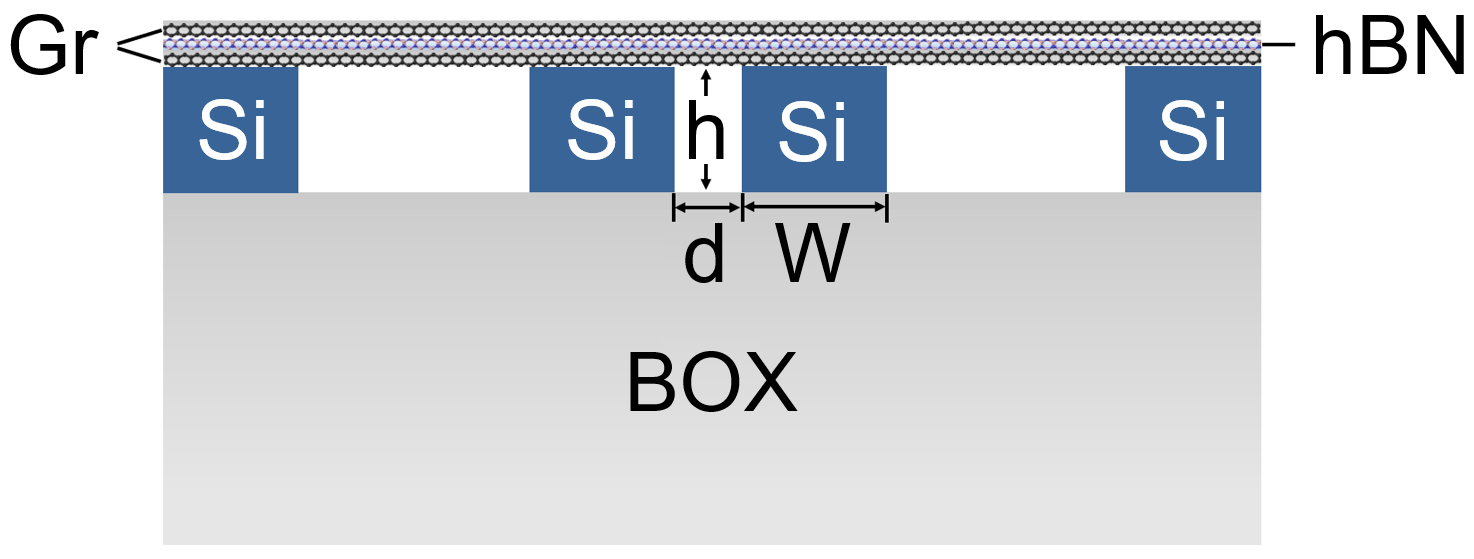}}
    \hfill
  \subfloat[\label{1b}]{%
        \includegraphics[width=1\linewidth]{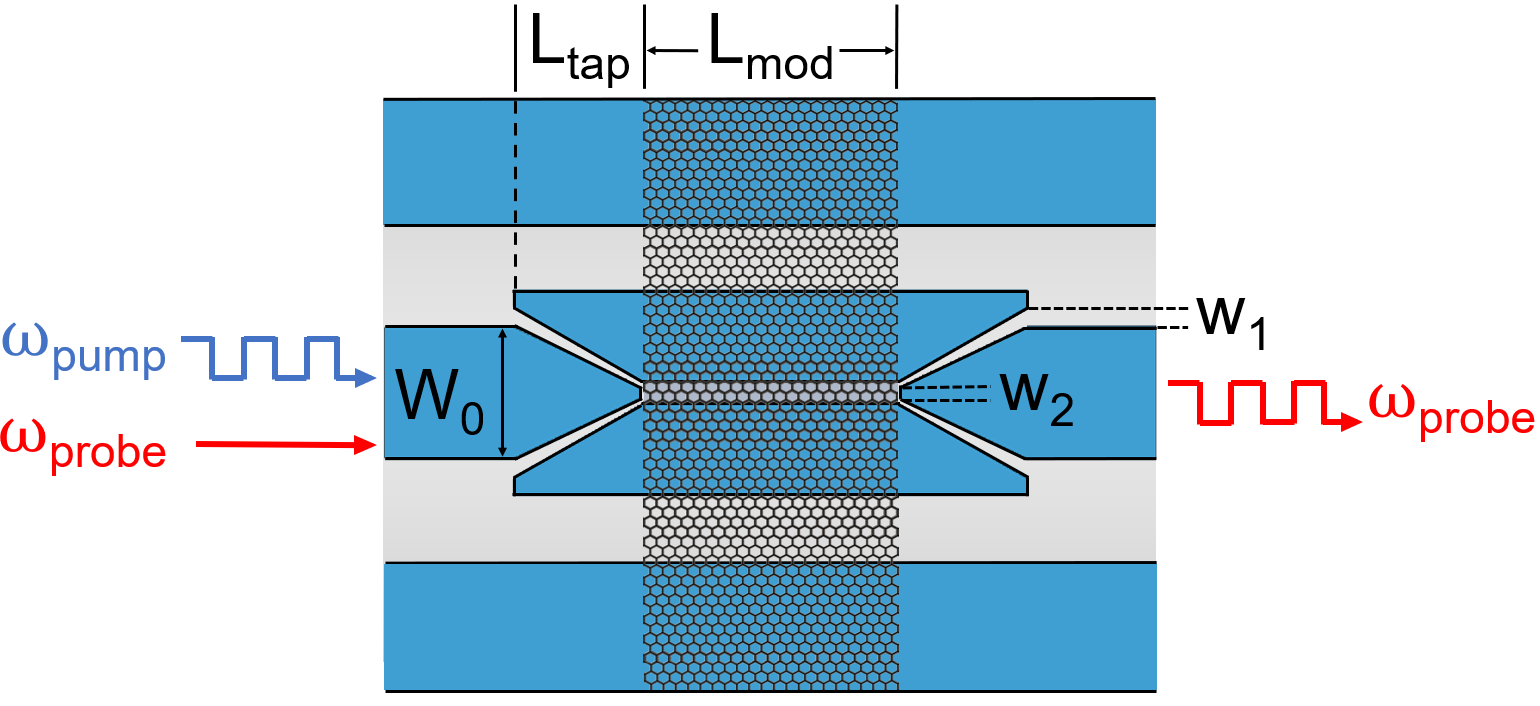}}
  \caption{(a) Front view and (b) top view of the on-chip modulator. Gr: graphene, Si: silicon, BOX: buried oxide, $h$: waveguide thickness, $d$: slot width, $W$: high-index region width, $W_0$: strip waveguide width, $w_1$: separation distance, $L_{\text{tap}}$: taper length, $L_{\text{mod}}$: modulator length, $w_2$: tip width. The pump signal modulates the probe signal. }
  \label{fig1} 
\end{figure}

\begin{figure} 
    \centering
  \subfloat[\label{slot_sweep}]{%
       \includegraphics[width=0.5\linewidth]{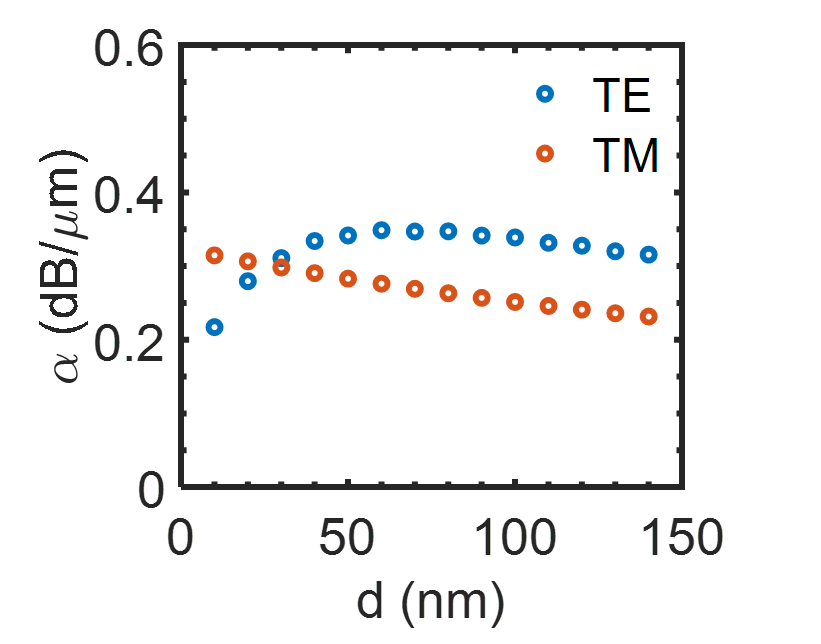}}
    \hfill
  \subfloat[\label{wavelength_sweep}]{%
        \includegraphics[width=0.5\linewidth]{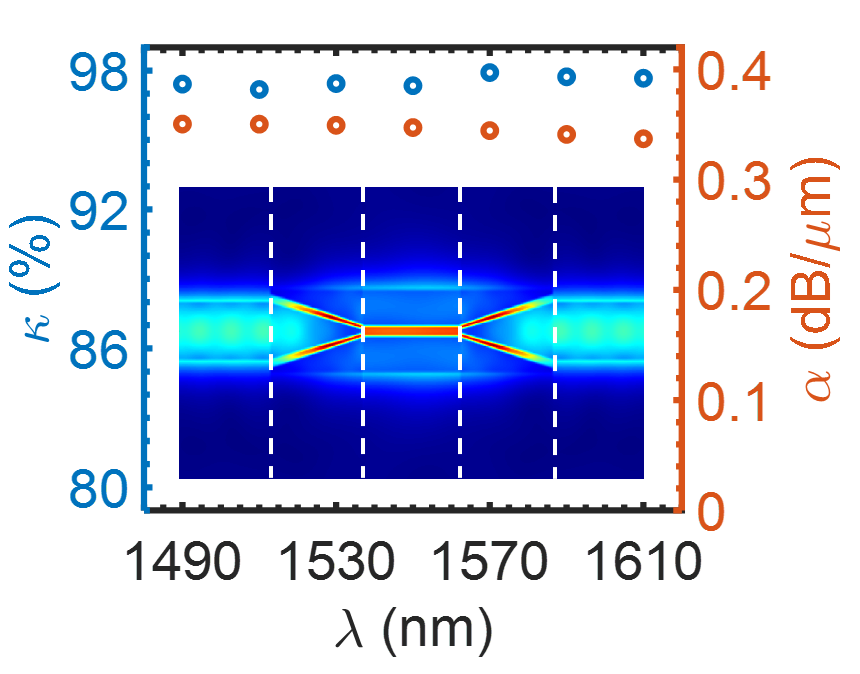}}
  \caption{(a) Propagation loss ($\alpha$) of the graphene-on-silicon slot waveguide as a function of the slot width ($d$) at $\lambda = 1550\,$nm. (b) Coupling efficiency ($\kappa$) and propagation loss ($\alpha$) as a function of wavelength ($\lambda$). The inset presents the eigenmode expansion (EME) simulation of the TE mode at $\lambda = 1550\,$nm.}
  \label{fig2} 
\end{figure}

\begin{figure} 
    \centering
  \subfloat[\label{length_absorption}]{%
       \includegraphics[width=0.5\linewidth]{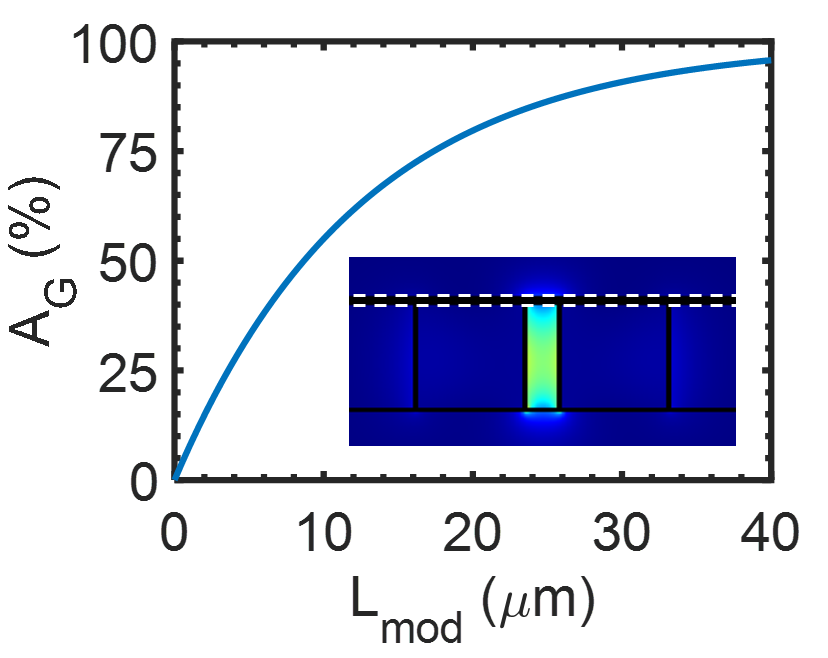}}
    \hfill
  \subfloat[\label{layers_absorption}]{%
        \includegraphics[width=0.5\linewidth]{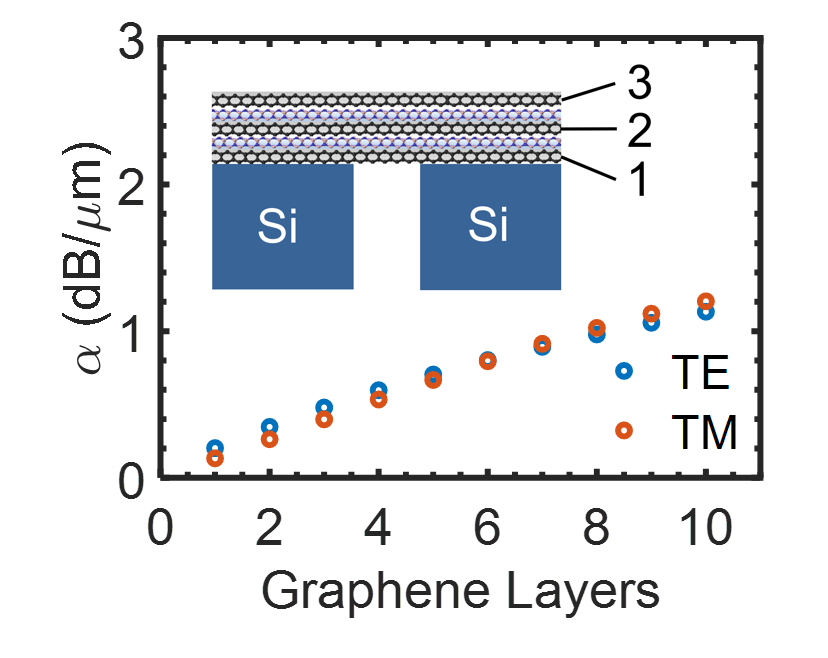}}
  \caption{(a) Percentage of power absorbed by graphene ($A_G$) as a function of the modulator length ($L_\text{mod}$) at $\lambda = 1550\,$nm. The inset shows the propagating TE mode in the slot waveguide. The dashed white lines represent the two graphene sheets. (b) Propagation loss ($\alpha$) as a function of the number of graphene layers at $\lambda = 1550\,$nm. The inset presents the numbering scheme of the graphene layers.}
  \label{fig3} 
\end{figure}

\begin{figure*}[t]
    \centering
    \includegraphics[width=0.8\textwidth]{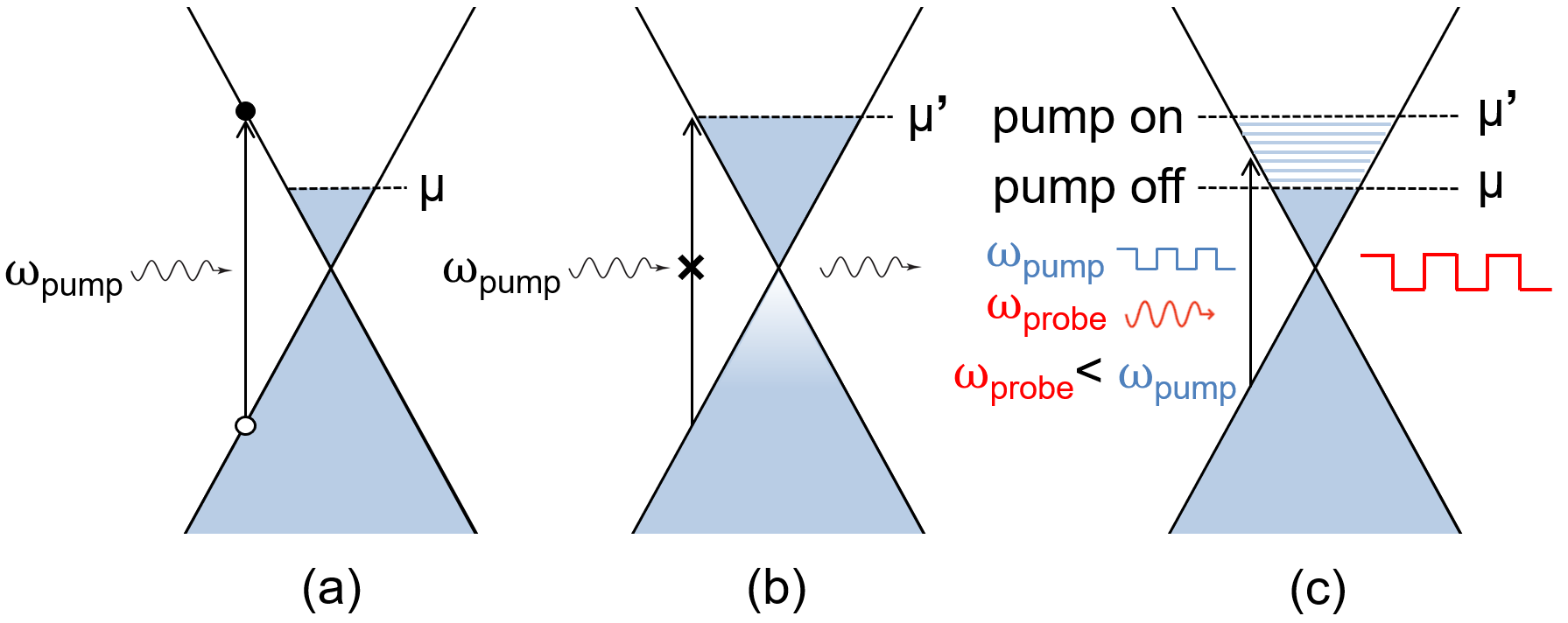}
    \caption{(a) Interband absorption of a pump photon with energy $\hbar \omega_{\text{pump}}$. (b) Pump photon Pauli-blocked (transmitted) after applying a sufficiently high pump intensity. (c) Transmission of the probe signal, as determined by the pump signal amplitude. Black and white circles represent electrons and holes, respectively. Filled energy states are represented by darker shades. Reprinted with permission \cite{alaloul2021low}. Copyright 2021, American Chemical Society.}
    \label{fig:transitions}
\end{figure*}

The structure of the on-chip modulator is illustrated in Fig. \ref{fig1}. It is composed of a graphene-hBN-graphene heterostructure that is placed on top of a silicon slot waveguide, which in turn sits atop a 3$\,$\textmu m thick buried oxide (BOX) layer. The dimensions of the device are given in Table \ref{my-label}. An 8$\,$\textmu m long taper is introduced before and after the modulator slot waveguide, to achieve a very high coupling efficiency ($\kappa$) of 97\% at 1550$\,$nm, which is in agreement with the experimental result that was reported in \cite{wang2009ultracompact} for a similar structure with the same taper length. The presence of the 2D heterostructure on top of the slot waveguide did not impact the coupling efficiency, because the simulated power overlap between the optical mode of the slot waveguide with and without it is 99.8\%. The slot width ($d$) was swept while the propagation loss of the waveguide ($\alpha$) was recorded, as shown in Fig. \ref{slot_sweep}. Because graphene is the only absorbing medium, a higher propagation loss corresponds to a higher graphene absorption. The 10$\,$nm hBN layer does not absorb light at telecom wavelengths, because hBN has an energy bandgap of $\sim6$$\,$eV \cite{elias2019direct}. As will be discussed in the next section, the modulation mechanism is based on the saturable absorption of graphene, so a higher graphene absorption is desirable. It is observed in Fig. \ref{slot_sweep} that the TE mode yields a higher absorption than the TM one, except at very small slot widths. The smaller the slot width, the more challenging it is to realize the device in practice because of limitations that are related to the minimum feature size in a fabrication process \cite{wang2016high}. Therefore, the TE mode is taken as the optimal mode for an $80\,$nm wide slot, because the absorption efficiency is very high at $d = 80\,$nm, and silicon slot waveguides with a similar width have been demonstrated in refs. \cite{cheng2015graphene, serna2014potential}. Fig. \ref{wavelength_sweep} shows the computed $\kappa$ and $\alpha$ as a function of wavelength, where it is observed that their variations in the studied wavelength range are minimal. These values are later used in section \ref{broadband} to characterize the broadband response of the device.

Fig. \ref{length_absorption} plots the percentage of power that is absorbed by graphene ($A_G$) as a function of $L_{\text{mod}}$, where $A_G$ is calculated from $\alpha$ using the Beer-Lambert Law:

\begin{equation} \label{abs}
    A_G (L_{\text{mod}}) = 1 - 10^{-(\alpha /10)*L_{\text{mod}}}
\end{equation}

The double layer graphene absorbs more than 95\% of the optical power when $L_{\text{mod}} = 40\,$\textmu m. Additional graphene-hBN layers may be stacked to boost the modulation efficiency of the device, because the absorption increases with the number of graphene layers (see Fig. \ref{layers_absorption}). For instance, the propagation loss exceeds 1$\,$dB/\textmu m for 10 graphene layers. However, each additional layer may require an additional step that involves exfoliation or chemical vapor deposition (CVD) \cite{deokar2015towards}, which might complicate the fabrication process. Therefore, the device operation is mainly studied for the case of double layer graphene, whereas the case of multiple graphene layers is briefly discussed. Alternatively, it is possible to scale the absorption by using multi-layered graphene, e.g. bilayer graphene \cite{liu2018graphene}. Nevertheless, the saturable absorption and the heating and cooling dynamics of multi-layered graphene are generally less efficient than in monolayer graphene, as will be discussed in sections \ref{modulation} \& \ref{performance}. Thus, this design is based on stacks of monolayer graphene and hBN to achieve a high modulation efficiency and an ultrafast response while maintaining a low $IL$.

\section{Operating Principle}
\label{principle}

\begin{figure} 
    \centering
  \subfloat[\label{fig:switch}]{%
       \includegraphics[width=0.5\linewidth]{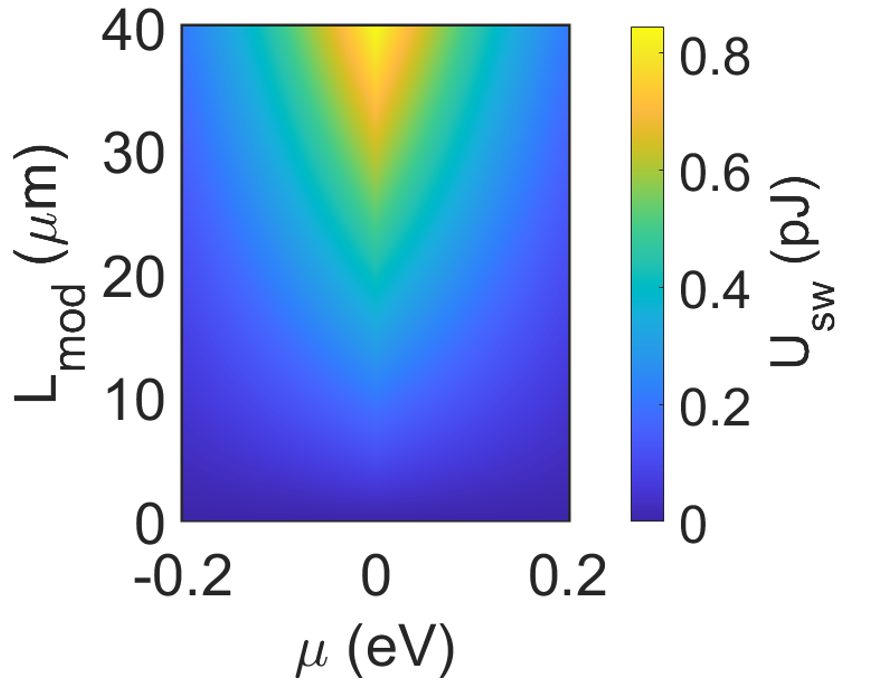}}
    \hfill 
  \subfloat[\label{fig:effective}]{%
        \includegraphics[width=0.5\linewidth]{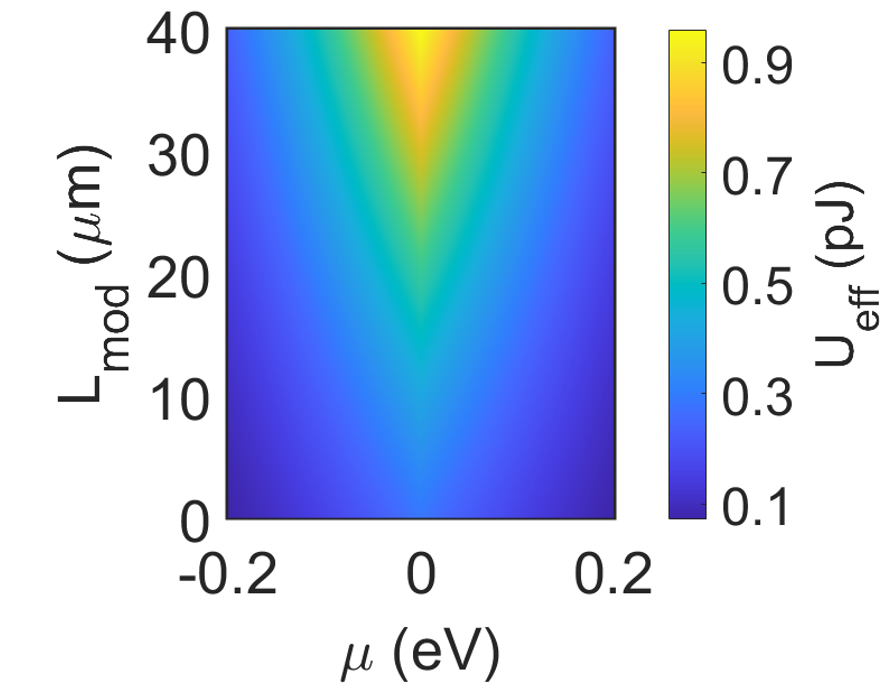}}
  \caption{(a) Switching energy ($U_{\text{sw}}$), and (b) effective switching energy ($U_{\text{eff}}$) as a function of chemical potential and modulator length. $\lambda_{\text{pump}}=$ 1550$\,$nm.}
  \label{newfig1} 
\end{figure}

The absorption of graphene can be tuned based on the principle of Pauli-blocking \cite{li2014ultrafast, chen2015all, liao2018ultra}, which occurs when photoexited electrons fill the conduction band states of graphene following a sufficiently intense pump excitation, thereby blocking the interband transition of other electrons \cite{alaloul2021low}. Figure \ref{fig:transitions}a illustrates the interband absorption of a pump photon with an energy $\hbar\omega_{\text{pump}}>2|\mu|$, where $\mu$ is the chemical potential of graphene. Upon absorbing a sufficiently intense pump excitation, electrons fill up the conduction band states, leading to a greater chemical potential $\mu^\prime$. As a consequence, incoming pump photons cannot induce an interband transition because $\hbar\omega_{\text{pump}} < 2|\mu^\prime|$ \cite{liu2018graphene}, and are thus transmitted through graphene (see Fig. \ref{fig:transitions}b). Similarly, a probe photon with an energy $\hbar\omega_{\text{probe}} \leq \hbar\omega_{\text{pump}}$ is also transmitted because $\hbar\omega_{\text{probe}} < 2|\mu^\prime|$. By utilizing this phenomenon, all-optical modulation is realized: the probe signal is transmitted when the pump signal is HIGH, or absorbed when the pump signal is LOW \cite{alaloul2021low}. Hence, the modulation mechanism is based on the saturable absorption of graphene \cite{li2014ultrafast, chen2015all, liao2018ultra}. From Fig. \ref{fig:transitions}b, it can be inferred that $\mu^\prime \approx \hbar\omega_{\text{pump}}/2$ \cite{ono2020ultrafast,alaloul2021low}. Then, the increase in carrier density ($\Delta n$) that is needed to saturate the absorption of graphene is given by \cite{alaloul2021low}:

\begin{equation}
    \Delta n = \dfrac{1}{\pi} \left(\dfrac{\Delta \mu}{\hbar v_{F}} \right)^2
\end{equation}

\noindent where $\hbar$ is the reduced Planck constant, $v_{F}$ is the Fermi velocity, and $\Delta \mu = \mu^\prime - \mu$ is the increase in chemical potential that is needed to induce Pauli-blocking. For a graphene sheet with an area $A=WL$, the number of carriers that are needed to reach $\mu^\prime$ is $m = \Delta n W L$. For the modulator device that is presented in Fig. \ref{fig2}, $L$ is taken as $L_{\text{mod}}$, and $W$ is the effective absorbing width (see supplementary section S1). Because each absorbed photon generates an electron-hole pair, the energy that is required to saturate the absorption of graphene ($U_{\text{sw}}$) is expressed as \cite{ono2020ultrafast, alaloul2021low}:

\begin{equation} \label{switchingEqation}
U_{\text{sw}} = \sum_{m} \hbar\omega_{m}
\end{equation}

\noindent where $U_\text{sw}$ is the switching energy, and $\hbar \omega$ is the pump photon energy. Fig. \ref{fig:switch} presents the switching energy ($U_{\text{sw}}$) for $\lambda_{\text{pump}} = 1550\,$nm. It is observed that $U_{\text{sw}}$ is higher at low chemical potentials and is maximized when the charge neutrality point coincides with the Dirac point ($\mu = 0$). That is because more electrons are required to fill the conduction band states up to $\mu^\prime$ when $\mu$ is low. Moreover, $U_{\text{sw}}$ increases with $L_{\text{mod}}$ because the larger the area, the higher is the switching energy according to $m = \Delta n W L$ and Eq. \ref{switchingEqation}. In practice, graphene may not fully absorb the waveguide mode. In addition, part of the optical mode energy is lost because of the waveguide's $IL$. To account for that, the effective switching energy ($U_{\text{eff}}$) is calculated as \cite{alaloul2021low}:

\begin{equation} \label{ueff_energy}
    U_{\text{eff}} = \dfrac{U_{\text{sw}}(1 + \Gamma + A_{\text{WG}})}{A_{\text{G}}*{(1-A_{\text{ns}})}}
\end{equation}

\noindent where $\Gamma = 1-\kappa$ is the coupling loss, $A_{\text{ns}}$ is the non-saturable fraction of $A_G$ \cite{bao2009atomic, bao2011monolayer, zhang2015dependence}, and is taken as 5\% for monolayer graphene \cite{alaloul2021low, bao2011monolayer}, and $A_{\text{WG}}$ is the slot waveguide loss that is not related to graphene. A 7$\,$dB/cm propagation loss has been reported in \cite{serna2014potential}, for a silicon slot waveguide with an 80$\,$nm wide slot. This propagation loss is 3 orders of magnitude lower than the propagation loss that is induced by double layer graphene ($\sim 0.35\,$dB/\textmu m). The calculated $U_{\text{eff}}$ is presented in Fig. \ref{fig:effective} for $\lambda_{\text{pump}} = 1550\,$nm. It is observed that $U_{\text{eff}}$ is higher than $U_{\text{sw}}$ and shares a similar trend with respect to $\mu$ and $L_{\text{mod}}$. Graphene is unintentionally doped when placed on a substrate, yielding chemical potentials in the range of 0.1--$\,$0.2$\:$eV \cite{ooi2017all, alaloul2021plasmon, romero2008n}. Within that range, $U_{\text{eff}}$ is $<543\,$fJ at all modulator lengths, and for a 20$\,$\textmu m long modulator, it is $<326\,$fJ.

\section{Modulation Efficiency}
\label{modulation}

The modulation efficiency of the modulator is quantified by its extinction ratio ($ER$) \cite{ono2020ultrafast, sun2019all}, which is given by:

\begin{equation} \label{ER_eq}
    ER = 10\,\text{log}_{10} \left( \dfrac{T_{\text{on}}}{T_{\text{off}}} \right)
\end{equation}

\noindent where $T_{\text{on}}$ and $T_{\text{off}}$ represent the transmitted power of the probe signal when the pump signal is turned on and off, respectively. The absorption of graphene is maximized when the pump signal is turned off, whereas the maximum transmission of the probe signal ($T_{\text{max}}$) is obtained when a pump signal with an energy $U>U_{\text{eff}}$ is applied. $T_{\text{max}}$ and $T_{\text{off}}$ can be expressed as \cite{alaloul2021low}:

\begin{equation} \label{Tmax}
    T_{\text{max}} = [1 - (\Gamma + A_{\text{WG}} + A_{\text{Gr}}A_{\text{ns}})]* (1-\Gamma)
\end{equation}

\begin{equation} \label{Toff}
    T_{\text{off}} = [1 - (\Gamma + A_{\text{WG}} + A_{\text{Gr}})] * (1-\Gamma)
\end{equation}

The $IL$ is quantified by the following relation \cite{FONG200212-1}:

\begin{equation} \label{IL_eq}
    IL = 10\,\text{log}_{10}\left(\dfrac{1}{T_{\text{on}}} \right)
\end{equation}

\begin{figure} 
    \centering
  \subfloat[\label{ERIL}]{%
       \includegraphics[width=0.5\linewidth]{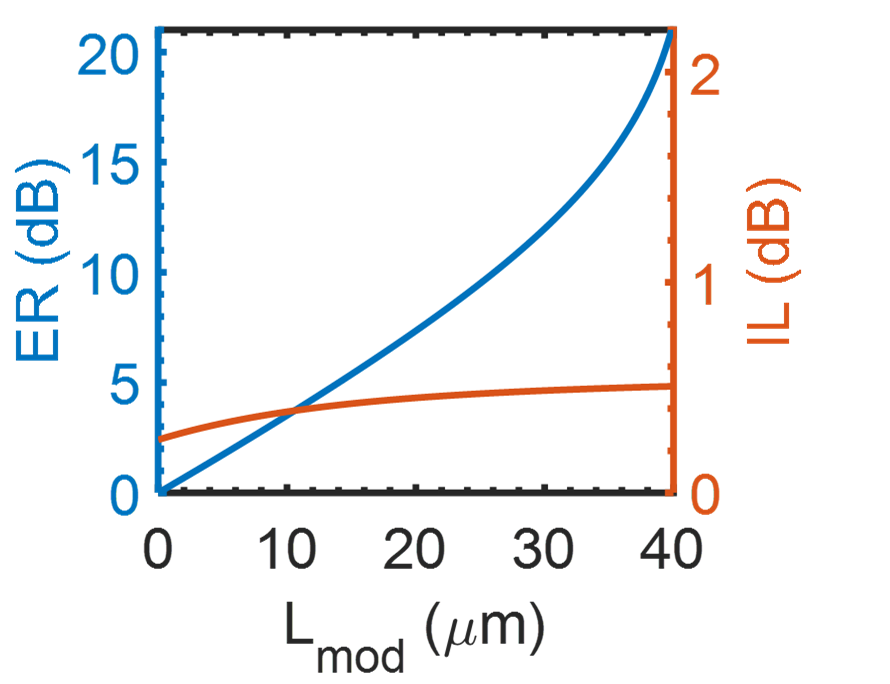}}
    \hfill
  \subfloat[\label{ER_layers}]{%
        \includegraphics[width=0.5\linewidth]{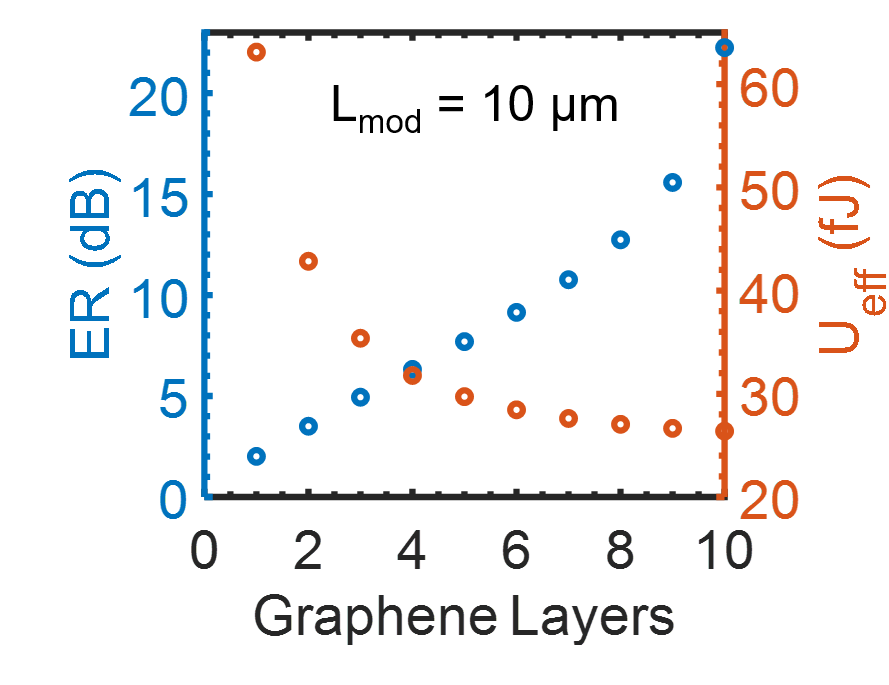}}
  \caption{(a) Maximum extinction ratio ($ER$) and insertion loss ($IL$) as a function of the modulator length ($L_\text{mod}$) for the double layer device at $\lambda_{\text{probe}} = 1550\,$nm. (b) Maximum extinction ratio ($ER$) and effective switching energy ($U_{\text{eff}}$) as a function of the number of graphene layers for a 10$\,$\textmu m long modulator at $\lambda_{\text{probe}} = 1550\,$nm.}
  \label{ER} 
\end{figure}

Figure \ref{ERIL} presents the maximum $ER$ and the $IL$ as a function of $L_{\text{mod}}$ for the double layer device. It is observed that $ER$ increases with $L_{\text{mod}}$ and exceeds 20$\,$dB for a 40$\,$\textmu m long modulator. For a 20$\,$\textmu m long modulator, the corresponding $ER$ is 7.3$\,$dB. It is noted that the $IL$ is ultra-low ($<0.6\,$dB) at all modulator lengths. Figure \ref{ER_layers} presents the $ER$ and $U_{\text{eff}}$ as a function of the number of graphene layers for a 10$\,$\textmu m long modulator. It is observed that the $ER$ increases as the number of graphene layers increases, whereas $U_{\text{eff}}$ decreases, which is consistent with Eqs. \ref{abs}, \ref{ueff_energy}, \ref{ER_eq}, \ref{Tmax} \& \ref{Toff}, because graphene's propagation loss increases with the number of graphene layers (see Fig. \ref{layers_absorption}). As explained in supplementary section S2, a portion of the optical mode propagates in the upper graphene and hBN layers, which increases the coupling loss and the overall propagation loss of the optical mode.

For monolayer graphene, the non-saturable absorption ($A_{\text{ns}}$) can be as low as 5\%, as was previously mentioned. For multilayer graphene, $A_{\text{ns}}$ is higher, which results in a higher $IL$ according to Eqs. \ref{Tmax}, \ref{Toff} \& \ref{IL_eq}. For instance, for multi-layer graphene with $<5$ layers, the reported $A_{\text{ns}}$ is $(31.48/74.17)*100\% = 42\%$, and for graphene with $<9$ layers, the reported $A_{\text{ns}}$ is $(67.85/86.02)*100\% = 78.88\%$ \cite{bao2011monolayer}. Therefore, even though multi-layer graphene exhibits a higher optical absorption than monolayer graphene, its saturable absorption is less efficient, which leads to a higher $IL$. Hence, in this design, multiple layers of monolayer graphene with sandwiched hBN layers are stacked to achieve efficient saturable absorption and a low $IL$.

\section{Broadband Response}
\label{broadband}

\begin{figure} 
    \centering
  \subfloat[\label{probe}]{%
       \includegraphics[width=0.5\linewidth]{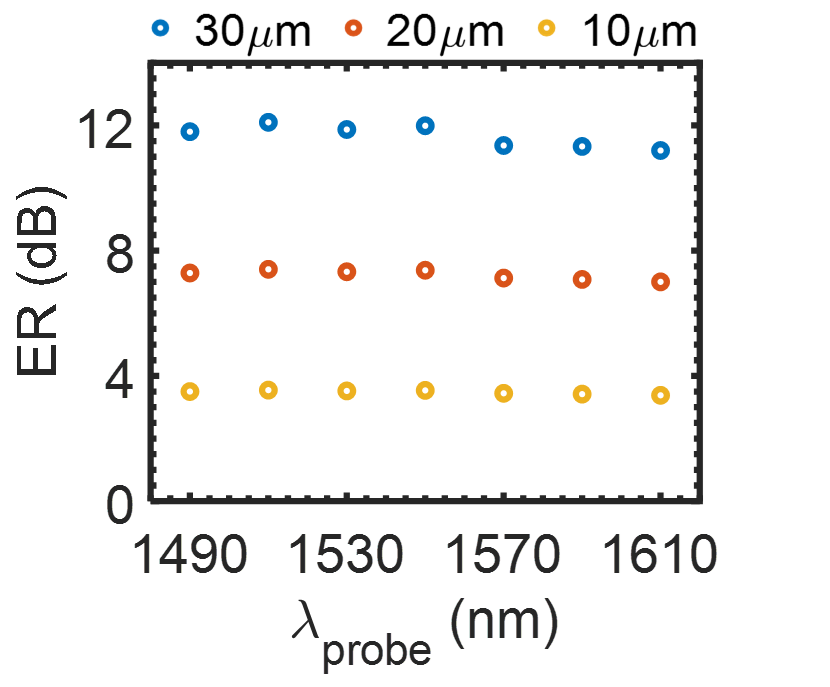}}
    \hfill
  \subfloat[\label{pump}]{%
        \includegraphics[width=0.5\linewidth]{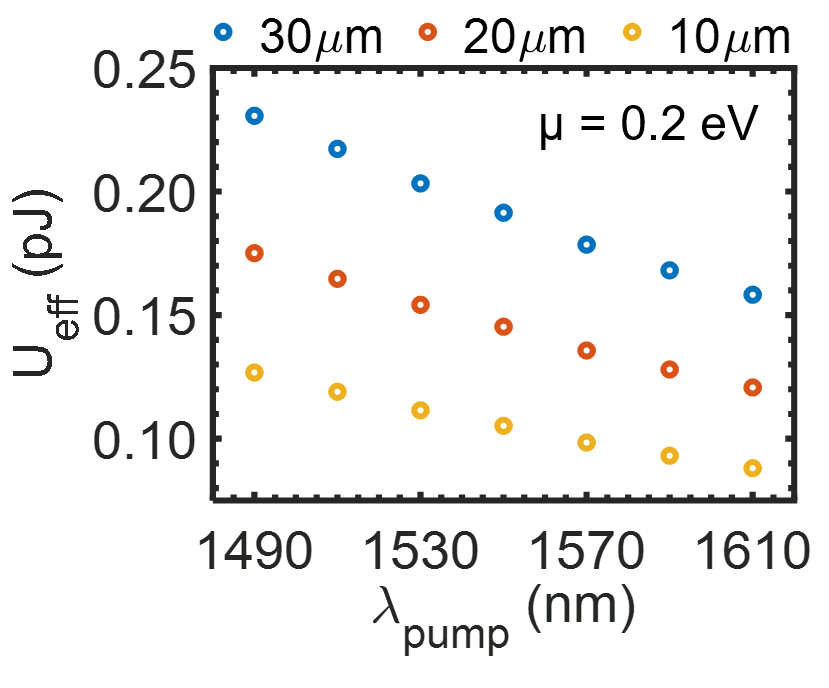}}
  \caption{(a) Maximum extinction ratio ($ER$) for three modulator lengths: 10$\,$\textmu m, 20$\,$\textmu m and 30$\,$\textmu m as a function of the probe signal wavelength ($\lambda_{\text{probe}}$), and (b) the effective switching energy ($U_{\text{eff}}$) as a function of the pump signal wavelength ($\lambda_{\text{pump}}$) for the same lengths and for $\mu = 0.2\,$eV.}
  \label{broadband_response} 
\end{figure}

The broadband response of the device is studied by calculating the $ER$ and $U_{\text{eff}}$ at other wavelengths. Fig. \ref{probe} presents the $ER$ as a function of the probe signal wavelength ($\lambda_{\text{probe}}$), and Fig. \ref{pump} presents $U_{\text{eff}}$ as a function of the pump signal wavelength ($\lambda_{\text{pump}}$) for $\mu = 0.2\,$eV. It is observed that the $ER$ does not vary much in the studied wavelength range, which is expected because the variations of $\kappa$ and $\alpha$ in this range are minimal, as was shown in Fig. \ref{wavelength_sweep}. Unlike the $ER$, $U_{\text{eff}}$ exhibits a relatively strong dependence on $\lambda_{\text{pump}}$, where it is higher at shorter wavelengths. This agrees with experimental observations \cite{zhang2015dependence, demetriou2016nonlinear}, where it has been reported that a lower pump intensity is required to saturate graphene at longer pump wavelengths. This can be explained by the unique conical dispersion of graphene, which increases and widens at higher chemical potentials. Therefore, a pump signal with a higher energy would be required at short wavelengths to satisfy the Pauli-blocking condition, which is given by $\mu^\prime \approx \hbar\omega_{\text{pump}}/2$, as was explained in section \ref{principle}. $U_{\text{eff}}$ values at other modulator lengths and chemical potentials are presented in supplementary section S3.

\section{Modulation Performance}

\label{performance}

The device performance is determined by the electron heating and cooling dynamics in graphene \cite{alaloul2021low,li2014ultrafast}. Graphene has a unique conical dispersion, where its density of states fades away at the Dirac point. Consequently, electrons in the vicinity of the Dirac point have a relatively low heat capacity. Upon photoexcitation, these electrons immediately scatter with one another, creating an ephemeral Fermi-Dirac distribution of hot thermalized electrons within a few tens to 150 femtoseconds \cite{tielrooij2015generation, alaloul2021plasmon, chen2019highly}. Hence, photon energy is converted to electron heat. Eventually, hot electrons cool down in a few picoseconds by emitting optical and acoustic phonons, coupling with surface optical phonons, and most importantly through disorder-assisted scattering which dominates at room temperature \cite{song2012disorder, ma2014competing, chen2008charged, graham2013photocurrent}. Thus, electrons first heat up through intraband electron-electron scattering, then they cool down through phonon- and disorder-assisted scattering. In addition, fewer layer graphene is characterized by faster electron heating and cooling dynamics \cite{bao2011monolayer, alaloul2021low, newson2009ultrafast}.

In \cite{wang2016high}, a high-responsivity graphene-on-silicon slot waveguide photodetector has been demonstrated. Nonetheless, the performance of this device was limited by its $\sim 50\,$\textmu s rise time, which would correspond to a 7$\,$KHz electrical bandwidth based on $BW_e = 0.35/t_r$, where $BW_e$ is the electrical bandwidth, and $t_r$ is the rise time \cite{brown1992bandwidth}. In the literature, waveguide-integrated graphene photodetectors have been demonstrated with a bandwidth exceeding 110$\,$GHz \cite{ding2020ultra, schall2018record, ma2018plasmonically}. The relatively low bandwidth of the aforementioned graphene-on-silicon slot waveguide photodetector was attributed to the slow carrier heating and cooling dynamics in suspended graphene, where substrate-induced disorder-assisted scattering is absent. In another work \cite{schuler2016controlled}, a graphene-on-silicon slot waveguide photodetector has been demonstrated, but with a 10$\,$nm hBN layer placed beneath the graphene sheet. The reported bandwidth for this device was 65$\,$GHz, even though its structure is not much different from the one reported in \cite{wang2016high}. Graphene is not suspended when it forms a Van der Waals (vdW) heterostrcuture with hBN. The electron cooling dynamics in graphene–hBN vdW heterostructures are dominated by hot electron coupling to hyperbolic phonon polaritons (HPhPs), not disorder-assisted scattering \cite{alaloul2021plasmon,tielrooij2018out, principi2017super, huang2020ultra}. Experimentally, graphene on hBN demonstrated faster electron cooling dynamics than graphene on Si/SiO$_2$ substrates \cite{tielrooij2018out, golla2017ultrafast}. Therefore, the cooling pathway introduced by HPhPs is more efficient than disorder-assisted scattering, which could explain the enhanced bandwidth of the device reported in \cite{schuler2016controlled}.

\begin{figure}
  \centering
  \includegraphics[width=0.5\linewidth]{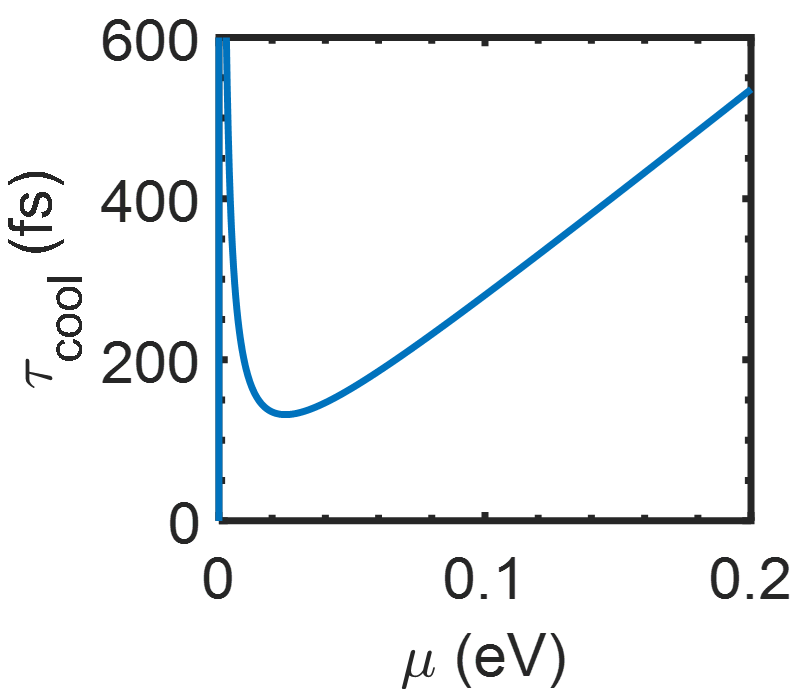}
\caption{Electron cooling time ($\tau_{\text{cool}}$) as a function of chemical potential ($\mu$).}
\label{fig:time}
\end{figure}

\begin{table*}
\label{tab10}
 \caption{Performance metrics of on-chip all-optical graphene switches/modulators}
\label{my-label1}
\begin{tabularx}{\textwidth}{@{}l*{8}{C}c@{}}
\toprule

Reference & $IL$ & $ER$ & $L$ & $ER$/$\mu$m & $IL$/$\mu$m  & $\tau_{\text{cool}}$ & $U_{\text{eff}}$ \\ 
\midrule

$ \:\:\:\:$ \cite{ono2020ultrafast} & 19$\,$dB & 3.5$\,$dB & 4$\,\text{\textmu}$m & 0.875$\,$dB/$\text{\textmu}$m & 4.75$\,$dB/$\text{\textmu}$m  & 260$\,$fs & 35$\,$fJ  \\ 

$ \:\:\:\:$ \cite{sun2019all} & n/a & 2.1$\,$dB & 10$\,\text{\textmu}$m & 0.21$\,$dB/$\text{\textmu}$m & n/a & n/a & n/a \textsuperscript{\emph{a}} \\

$ \:\:\:\:$ \cite{sun2018all} & negligible & 2.75$\,$dB & 100$\,\text{\textmu}$m & 0.0275$\,$dB/$\text{\textmu}$m & negligible & n/a  & n/a \textsuperscript{\emph{b}}  \\ 

$ \:\:\:\:$ \cite{wang2020cmos} & negligible & 1.1$\,$dB \textsuperscript{\emph{c}} & 30$\,\text{\textmu}$m & 0.0367$\,$dB/$\text{\textmu}$m & negligible & 1.2$\,$ps \textsuperscript{\emph{d}} & 1.38$\,$pJ \textsuperscript{\emph{e}} \\ 

$ \:\:\:\:$ \cite{qiu2021high} & $\sim 1\,$dB \textsuperscript{\emph{f}} & 11$\,$dB & 288$\,\text{\textmu}$m \textsuperscript{\emph{g}} & 0.038$\,$dB/$\text{\textmu}$m & 0.0035$\,$dB/$\text{\textmu}$m & 1.29$\,$\textmu s & n/a \textsuperscript{\emph{h}} \\

This work & $0.45\,$dB & 7.3$\,$dB & 20$\,\text{\textmu}$m & 0.365$\,$dB/$\text{\textmu}$m & $0.023\,$$\,$dB/$\text{\textmu}$m &  <600$\,$fs &  $<326\,$fJ \\ 

\bottomrule
\end{tabularx}
\textsuperscript{\emph{a}} Maximum input light power is 46$\,$mW;
\textsuperscript{\emph{b}} Input light power is 60$\,$mW;
  \textsuperscript{\emph{c}} Modulation depth is 22.7\%;
  \textsuperscript{\emph{d}} Limited by the resolution time of the asynchronous pump–probe system;
  \textsuperscript{\emph{e}} Saturation threshold;
  \textsuperscript{\emph{f}} Waveguide loss before transferring graphene.
  \textsuperscript{\emph{g}} Length of the graphene coating on the waveguide.
  \textsuperscript{\emph{h}} 90--$\,$109.6$\:$mW switching power.
  n/a: not available (not reported).
\end{table*}

An all-optical modulator that is based on graphene-on-silicon slot waveguides has not yet been demonstrated, to the best of our knowledge. As such, the electron heating and cooling dynamics and the corresponding response times of these devices are not reported in the literature. To ensure that our device performs ultrafast modulation and does not suffer from inefficient electron heating and cooling dynamics, hBN is sandwiched between the two graphene layers. In this heterostructure, the electron heat transfer channel is out-of-plane \cite{tielrooij2018out}, which means that the sandwiched hBN layer provides a cooling pathway for both the upper and the lower graphene layers.

In order to calculate the recovery time of the switch, the electrical conductivity of graphene is first calculated using the following relation \cite{lin2019asymmetric, alaloul2021plasmonic}:

\begin{equation} \label{eq:conductivity}
    \sigma = \sigma_{0}(1+\dfrac{\text{\textmu}^{2}}{\Delta^{2}}) \; , \; \;\sigma_{0} = 5(\dfrac{e^{2}}{h})
\end{equation}

\noindent where $h$ is Planck's constant. $\sigma_{0}$ is the minimum conductivity taken from \cite{lin2019asymmetric}, and $\Delta \approx 25\,$meV ($\lesssim 300\,$K) is a typical width of the charge neutrality region for graphene on a BN substrate \cite{song2011hot}. $\Delta$ is a fitting parameter that accounts for disorder in graphene, and it determines the mobility and the slope of the conductivity curve. Higher disorder is manifested in the form of a wider charge neutrality region in the conductivity plot \cite{chen2008charged, alaloul2021plasmon}. By fitting the conductivity curve using Eq. \ref{eq:conductivity}, it is possible to model graphene with different carrier mobilities. The electron cooling time is taken as the electron relaxation time. For the case of graphene-on-hBN, it can be calculated from $\sigma$ using the following relation \cite{hamidouche2020optoelectronic}:

\begin{equation}
    \tau_{\text{cool}} = \dfrac{2 \sigma}{e^2 v_F^2 D}  \; , \; \; D = \dfrac{g \mu}{2 \pi \hbar^2 v_F^2}
\end{equation}

\noindent where $e$ is the electron charge, $D$ is the density of states of graphene, and $g = 4$ is the total degeneracy due to the spin
degeneracy and the valley degeneracy \cite{liu2018graphene}. Fig. \ref{fig:time} presents the calculated $\tau_{\text{cool}}$ as a function of $\mu$, which is found out to be $<600\,$fs within the $\mu$ range of 0.1--$\,$0.2$\:$eV, where $\mu$ is related to the carrier density ($n$) by $\mu = \hbar v_F \sqrt{n}$ \cite{liu2018graphene}. This result is similar to the experimentally reported electron cooling times for graphene-on-hBN, which were in the range of 200--$\,$400$\:$fs in \cite{golla2017ultrafast}. The switching time of the device is on the order of the electron heating time ($<150\,$fs) \cite{tielrooij2015generation, alaloul2021plasmon, chen2019highly}, because this is the time in which a sea of hot electrons is induced following a pump excitation; these electrons fill up the conduction band states leading to Pauli-blocking \cite{alaloul2022electrical}. Then, the modulator reverts to its steady-state in a timescale of $\sim$$\,\tau_{\text{cool}}$, which is $<600\,$fs in the case that is considered in Fig. \ref{fig:time}. Based on these values, this device can theoretically operate as an all-optical modulator with a bandwidth in the hundreds of GHz \cite{li2014ultrafast}. The heating and cooling dynamics of additional graphene-hBN layers are expected to be similar to the case of double layer graphene, because each graphene layer forms a vdW heterostructure with hBN. \\ \indent The pump fluence affects the carrier cooling time. In \cite{golla2017ultrafast}, pump-probe spectroscopy of graphene-hBN heterostructures was performed with pump fluences of 80$\,\text{\textmu} \text{J/cm}^2$, 60$\,\text{\textmu} \text{J/cm}^2$, and 50$\,\text{\textmu} \text{J/cm}^2$, for which the extracted carrier cooling times were 375$\,$fs, 250$\,$fs, and 200$\,$fs ($\pm 25 \,$fs), respectively. These cooling times are similar to the electron cooling times that are presented in Fig. \ref{fig:time}. Therefore, the energy of the pump signal does not need to be higher than $U_{\text{eff}}$, because the electron cooling time increases at higher pump fluences, which in turn decreases the switching performance of the device. The studied vdW heterostructure in \cite{golla2017ultrafast}, consisted of graphene on a hBN layer. In addition, it has been reported that the encapsulation of graphene in hBN induces ultrafast cooling pathways of hot electrons by HPhPs \cite{principi2017super, tielrooij2018out}. It is possible to design high-performance all-optical modulators by building Gr/hBN heterostructures or by encapsulating graphene in hBN. \\ \indent Table \ref{my-label1} presents the reported performance metrics of demonstrated on-chip all-optical graphene switches and modulators. The device that is presented in \cite{ono2020ultrafast}, is based on graphene-loaded plasmonic slot waveguides. It achieves the highest modulation efficiency (0.875$\,$dB/\textmu m) and the lowest switching energy ($35\,$fJ), but is limited by an excessive $IL$. To meet the demands of next-generation telecom and datacom networks, photonic devices with an $IL<5\,$dB are required \cite{giambra2021wafer}. Nevertheless, this device has made a significant contribution in all-optical switching because it has demonstrated an unprecedented recovery time of 260$\,$fs using graphene. The authors explained that this result could be because of the ultrafast carrier diffusion out of the 30$\,$nm narrow slot region. To the best of our knowledge, a dedicated study of the influence of the graphene sheet area on the carrier relaxation time has not been reported in the literature. It is not inconceivable that the ultrafast recovery time of $260\,$fs was brought about by the small graphene sheet area. In fact, pump-probe spectroscopy of waveguide-integrated germanium revealed that the lifetime of photoexcited carriers increases with the germanium length and width \cite{srinivasan2016extraction}. This could also be the case for waveguide-integrated graphene devices. Therefore, it might be possible to achieve ultrafast switching beyond a recovery time of $260\,$fs, by building Gr/hBN all-optical modulators with a small device area; the smaller area enables faster carrier diffusion across the device, and the use of Gr/hBN heterostructures achieves ultrafast carrier cooling that is four times faster than that on a SiO$_2$ substrate \cite{golla2017ultrafast}. Therefore, further investigation of the influence of the graphene sheet area on the carrier cooling dynamics is required. In practice, however, the absorption of graphene is small, which necessitates the need for longer devices. The use of Gr/hBN heterostructures accelerates the carrier cooling dynamics of graphene, even for large area sheets. For instance, a 200$\,$fs ($\pm 25 \,$fs) carrier cooling time at a pump fluence of $50\,\mu$J/cm$^2$ was reported in \cite{golla2017ultrafast}, for a Gr/hBN heterostructure that is tens of micrometres long. The non-plasmonic devices that were reported in \cite{sun2018all, wang2020cmos, qiu2021high} have a much lower $IL$ and can achieve high modulation efficiencies at large $L$, but their switching energies may exceed $1\,$pJ/bit at these lengths. Though this device is longer than other devices, it achieves a relatively high modulation efficiency (0.365$\,$dB/$\text{\textmu}$m), energy-efficient switching ($<326\,$fJ) and ultrafast recovery ($<600\,$fs) at an almost negligible $IL$ of 0.45$\,$dB.\\
\indent The fabrication process of Si sot waveguides is presented in \cite{wang2016high}, where the waveguide pattern is defined on a commercial silicon-on-insulator (SOI) substrate using electron beam lithography (EBL). Optoelectronic devices with a Gr/hBN/Gr heterostructure were fabricated in \cite{song2021deep}. This heterostructure is prepared by the dry transfer technique \cite{wang2013one}, where graphene and hBN layers are first separately exfoliated onto Si/SiO$_2$ substrates, and each layer is then picked up by a polymer stamp. In \cite{argentero2017unraveling}, hBN was picked up by a single layer graphene that is attached to a PMMA membrane. By picking up an additional layer of graphene, a Gr/hBN/Gr heterostructure is formed. In \cite{wang2020cmos}, a waveguide-integrated all-optical graphene modulator with partially suspended graphene was demonstrated. There, a PMMA/Gr film was transferred onto the waveguides and patterned by EBL to control the modulation depth of the device. The redundant graphene layer was etched off by reactive ion etching (RIE) using oxygen. PMMA was removed by soaking the chip in acetone. By following a similar approach, a PMMA film with a Gr/hBN/Gr heterostructure could be transferred onto the Si slot waveguides, to build the double-graphene modulator that is shown in Fig. \ref{1a}.\\
\indent The presence of surface defects in graphene, like wrinkles, induces a scattering loss, which increases the non-saturable fraction of the absorption ($A_{\text{ns}}$) \cite{bao2011monolayer}. This in turn degrades the modulation efficiency, increases the switching energy, and contributes to a higher $IL$. Further advances in graphene transfer methods will contribute to the development of more efficient and reliable graphene-based devices.

\section{Conclusion}

We propose a novel design of a high-performance all-optical modulator that is based on graphene and hexagonal boron nitride (hBN) heterostructures, that are hybrid integrated into silicon slot waveguides. Simulations were performed to optimize the design, where an 8$\,$\textmu m long taper is introduced to achieve a 97\% coupling efficiency between the slotted modulator waveguide and silicon strip waveguides. Moreover, the switching energy, modulation efficiency, broadband response, and modulation performance of the device were investigated. Using this device, a high extinction ratio (ER) of 7.3$\,$dB, an ultra-low insertion loss (IL) of $<$0.6$\,$dB, and energy-efficient switching ($<$0.33$\,$pJ/bit) are attainable for a 20$\,\text{\textmu}$m long modulator with double layer graphene. In addition, it performs ultrafast switching with a recovery time of $<600\,$fs, and could potentially be employed as a high-performance all-optical modulator with an ultra-high bandwidth in the hundreds of GHz. Furthermore, stacking additional layers of Gr/hBN heterostructures enhances the modulation efficiency of the device, while theoretically maintaining the same ultrafast response. The high-performance metrics of this device across a wide wavelength range are expected to serve the needs of next-generation optical computing systems.

\section*{Acknowledgment}

Support from the NYUAD Center for Cyber Security research
grant is gratefully acknowledged.


%





\ifCLASSOPTIONcaptionsoff
  \newpage
\fi





\bibliographystyle{IEEEtran}
\bibliography{Bibliography}

\vfill


\end{document}


\maketitle

\section{Effective Absorbing Width}

\begin{figure} [htbp]
  \centering
  \includegraphics[width=1\linewidth]{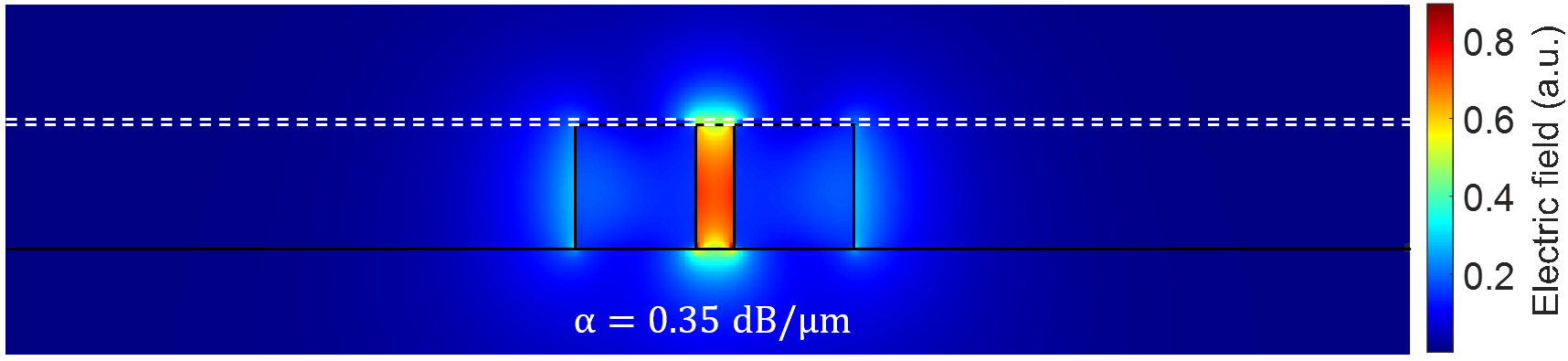}
\caption{Electric field profile of the TE mode for the double layer device with $3\,$\textmu m wide graphene and hBN layers. $\lambda = 1550\,$nm. The white dashed lines represent graphene.}
\label{fig:3micron}
\end{figure}

\begin{figure} [htbp]
  \centering
  \includegraphics[width=0.9\linewidth]{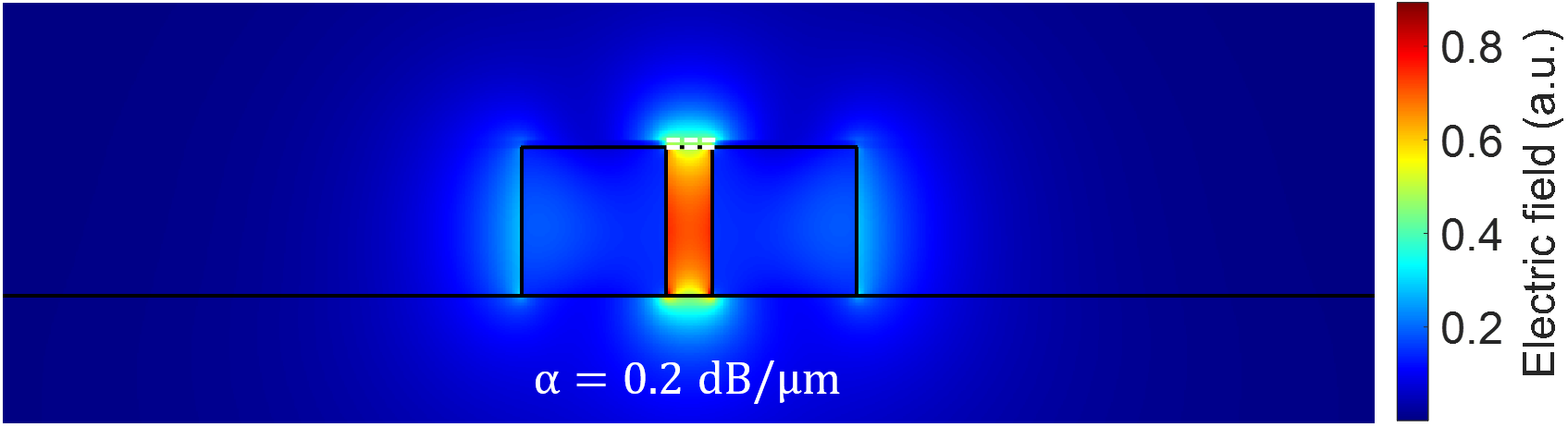}
\caption{Electric field profile of the TE mode for the double layer device with $80\,$nm wide graphene and hBN layers. $\lambda = 1550\,$nm. The white dashed lines represent graphene.}
\label{fig:80nano}
\end{figure}

\begin{figure} [htbp]
  \centering
  \includegraphics[width=0.9\linewidth]{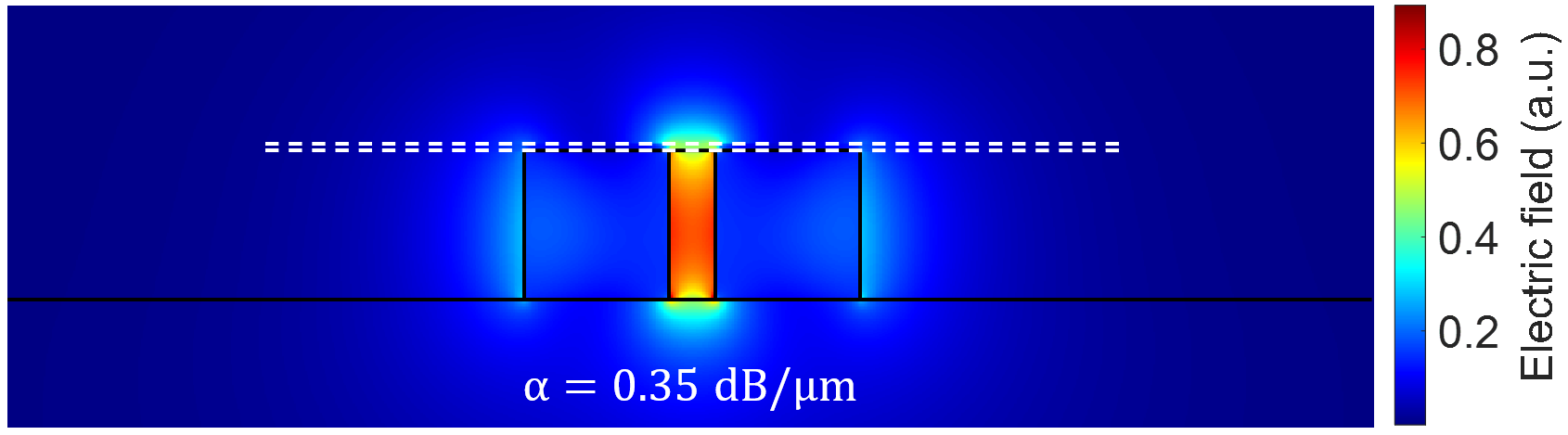}
\caption{Electric field profile of the TE mode for the double layer device with $1.4\,$\textmu m wide graphene and hBN layers. $\lambda = 1550\,$nm. The white dashed lines represent graphene.}
\label{fig:1400}
\end{figure}

The propagation loss ($\alpha$) for the quasi transverse-electric (quasi-TE) mode is $\sim 0.35\,$dB/\textmu m, as was shown in Fig. 2a of the main text. This $\alpha$ value was obtained by simulating a $3\,$\textmu m wide graphene and hexagonal boron nitride (hBN) layers on top of the silicon slot waveguide (see Fig. \ref{fig:3micron}). It might be assumed that the portion of graphene that is on top of the slotted region is the one that solely contributes to the absorption of the waveguide mode, because that is where the waveguide mode is mostly confined. However, when simulating the same structure but with 80$\,$nm wide graphene and hBN layers on top of the slotted region (see Fig. \ref{fig:80nano}), the resulting $\alpha$ is merely 0.2$\,$dB/\textmu m. This indicates that other portions of the graphene sheet significantly contribute to the absorption. Therefore, the width of the graphene and hBN layers is swept to find out the effective absorbing width which would yield a $\sim 0.35\,$dB/\textmu m propagation loss. It is found out that 1.4$\,$\textmu m wide graphene and hBN layers yield a $\sim 0.35\,$dB/\textmu m propagation loss (see Fig. \ref{fig:1400}). Hence, 1.4$\,$\textmu m is taken as the effective absorbing width for calculating the switching energy.

\section{Multiple Graphene Layers}

\begin{figure}
  \centering
  \includegraphics[width=0.5\linewidth]{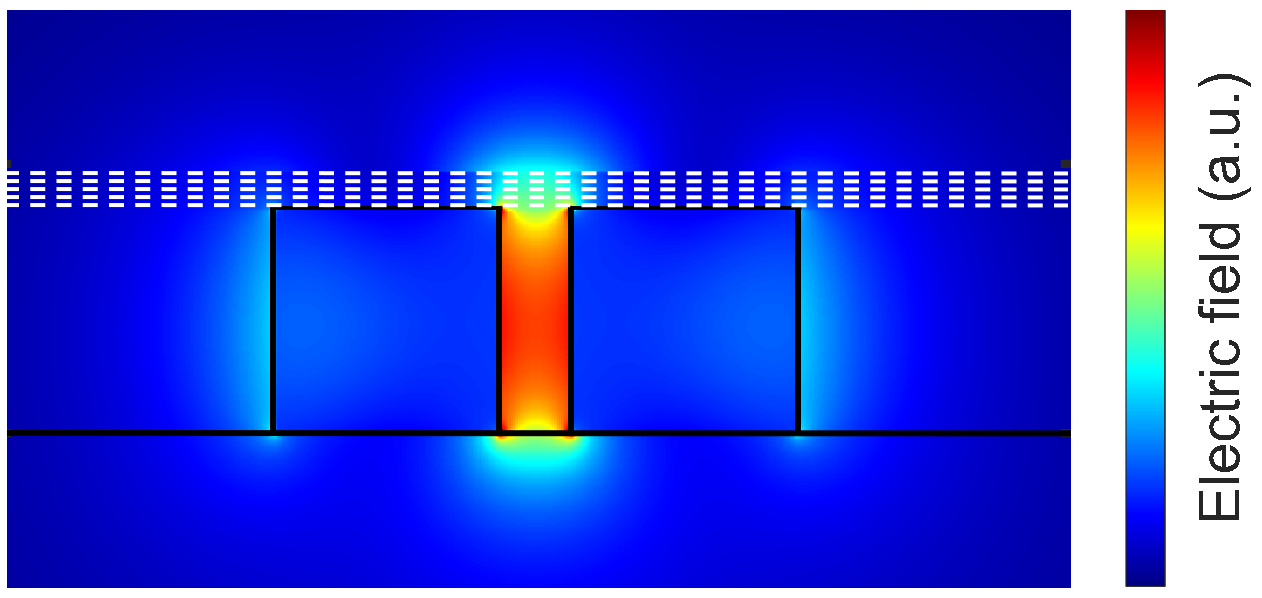}
\caption{The computed electric field profile of the TE mode for the device with five layers of graphene. $\lambda = 1550\,$nm. The white dashed lines represent graphene.}
\label{fig:layers}
\end{figure}

\begin{figure}
  \centering
  \includegraphics[width=0.45\linewidth]{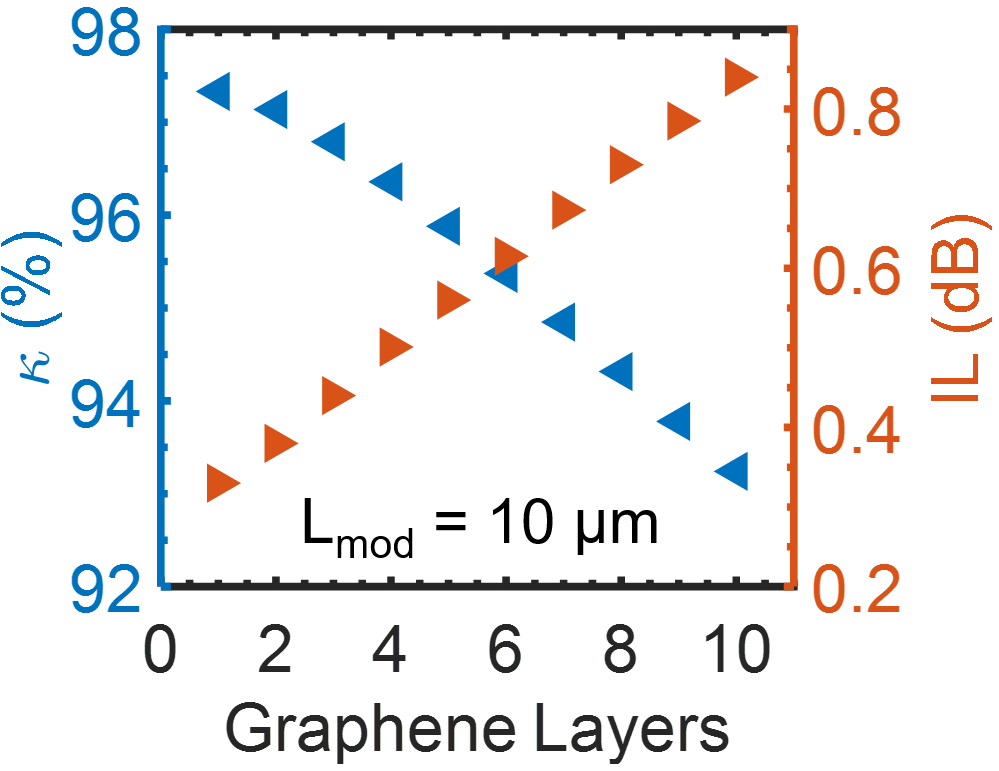}
\caption{Coupling efficiency ($\kappa$) of the device as a function of the number of graphene layers. $\lambda = 1550\,$nm. The resultant insertion loss ($IL$) as a function of the number of graphene layers. $\lambda_{\text{probe}} = 1550\,$nm and $L_{\text{mod}} = 10\,$\textmu m. }
\label{fig:IL2}
\end{figure}

When simulating the structure with multiple graphene and hBN layers, it is found out that a portion of the optical mode is guided in the Gr/hBN layers because they have a higher refractive index than air, so they concentrate part of the in-plane electric field, which interacts with graphene leading to a higher overall absorption. Therefore, there is a 2-step enhancement of the absorption for the case of multiple Gr/hBN layers. First, the guided optical mode is enhanced at the air/Si interface in the slotted region, which increases the effective absorption of graphene, then the absorption is also enhanced by the partial presence of the guided optical mode in the Gr/hBN layers. To illustrate this, Fig. \ref{fig:layers} shows the computed TE-mode with five layers of graphene. Furthermore, as the number of graphene and hBN layers increase, a greater portion of the optical mode is guided in these layers, which introduces a higher mode mismatch with the original strip waveguide mode. This in turn results in a higher coupling loss and a higher insertion loss ($IL$), as shown in Fig. \ref{fig:IL2}.

\section{Effective Switching Energy}

\begin{figure}
\begin{subfigure}{.5\textwidth}
  \centering
  \includegraphics[width=0.9\linewidth]{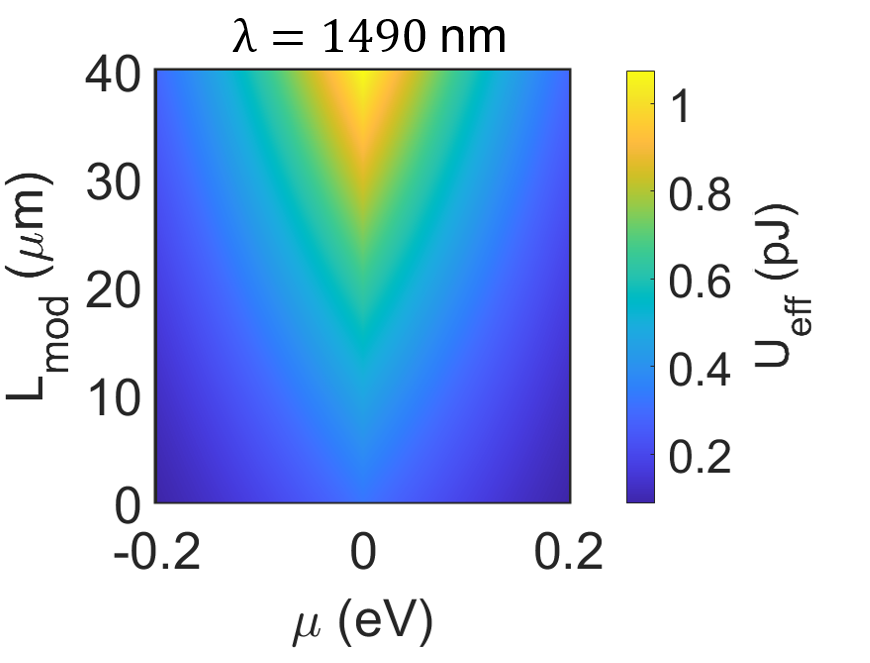}
  \caption{}
  \label{fig:1490}
\end{subfigure}%
\begin{subfigure}{.5\textwidth}
  \centering
  \includegraphics[width=0.9\linewidth]{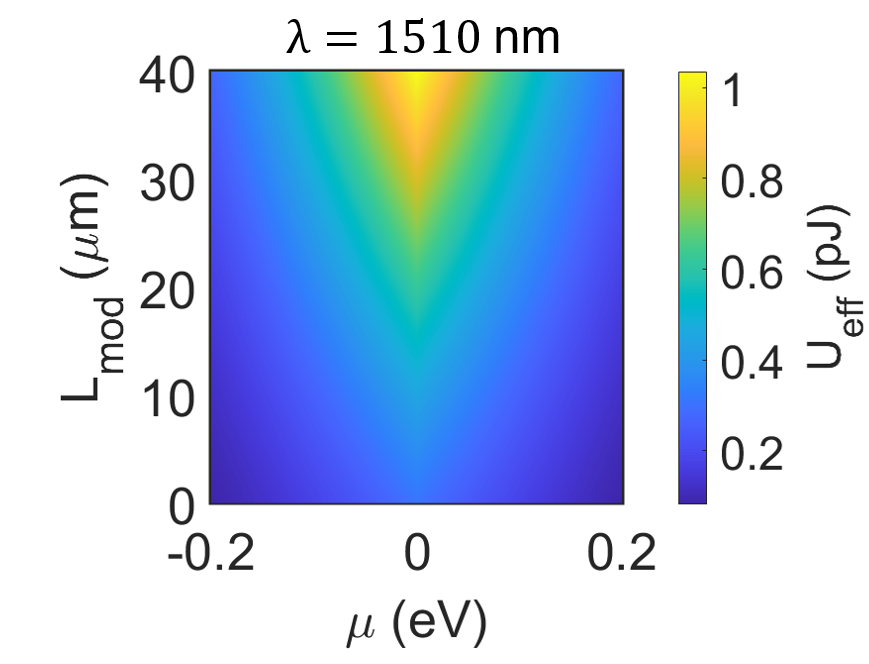}
  \caption{}
  \label{fig:1510}
\end{subfigure}
\caption{(a) Effective switching energy ($U_{\text{eff}}$) as a function of chemical potential ($\mu$) and modulator length ($L_{\text{mod}}$). $\lambda_{\text{pump}}=$ 1490$\,$nm. (b) $U_{\text{eff}}$ as a function of chemical potential and modulator length. $\lambda_{\text{pump}}=$ 1510$\,$nm.}
\label{fig:totITO}
\end{figure}

\begin{figure}
\begin{subfigure}{.5\textwidth}
  \centering
  \includegraphics[width=0.9\linewidth]{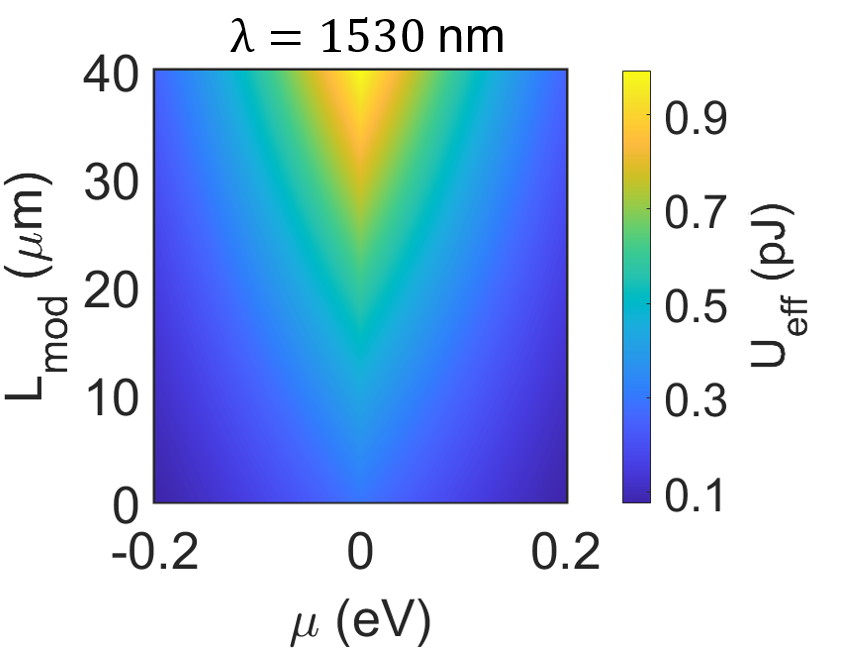}
  \caption{}
  \label{fig:1530}
\end{subfigure}%
\begin{subfigure}{.5\textwidth}
  \centering
  \includegraphics[width=0.9\linewidth]{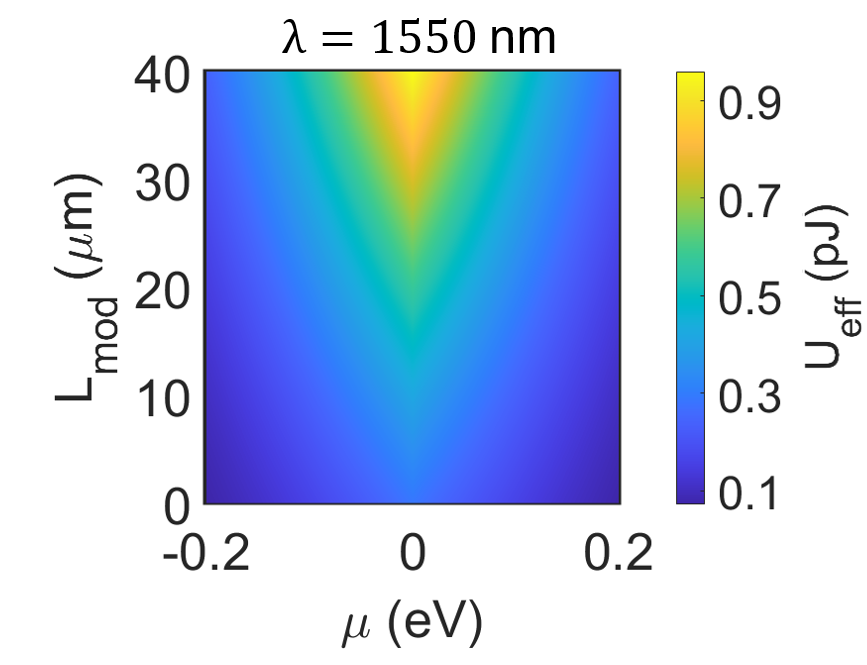}
  \caption{}
  \label{fig:1550}
\end{subfigure}
\caption{(a) Effective switching energy ($U_{\text{eff}}$) as a function of chemical potential ($\mu$) and modulator length ($L_{\text{mod}}$). $\lambda_{\text{pump}}=$ 1530$\,$nm. (b) $U_{\text{eff}}$ as a function of chemical potential and modulator length. $\lambda_{\text{pump}}=$ 1550$\,$nm.}
\label{fig:totT1O}
\end{figure}

\begin{figure}
\begin{subfigure}{.5\textwidth}
  \centering
  \includegraphics[width=0.9\linewidth]{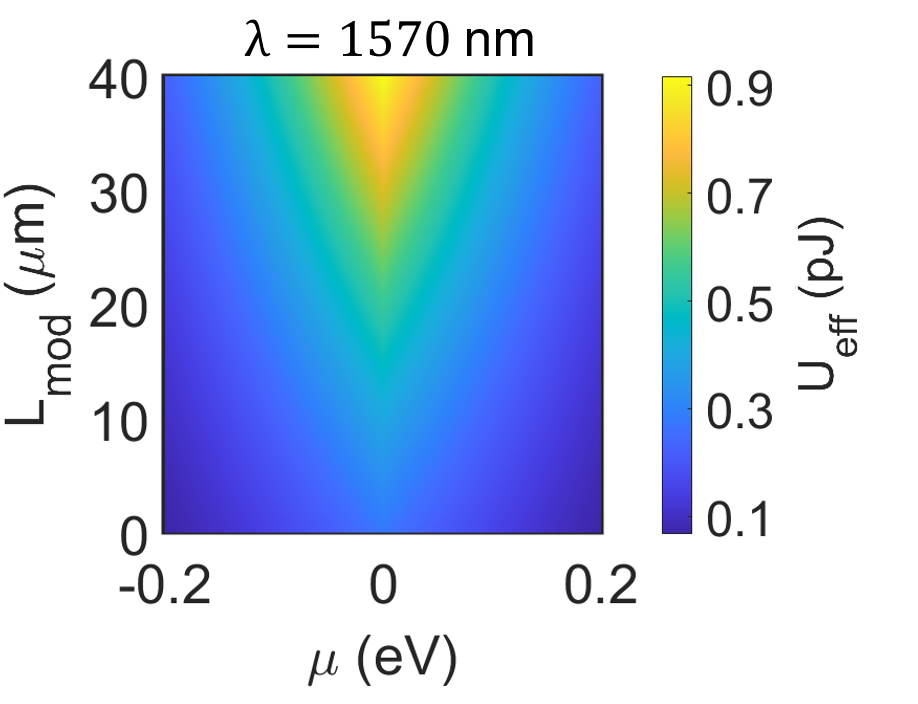}
  \caption{}
  \label{fig:1570}
\end{subfigure}%
\begin{subfigure}{.5\textwidth}
  \centering
  \includegraphics[width=0.9\linewidth]{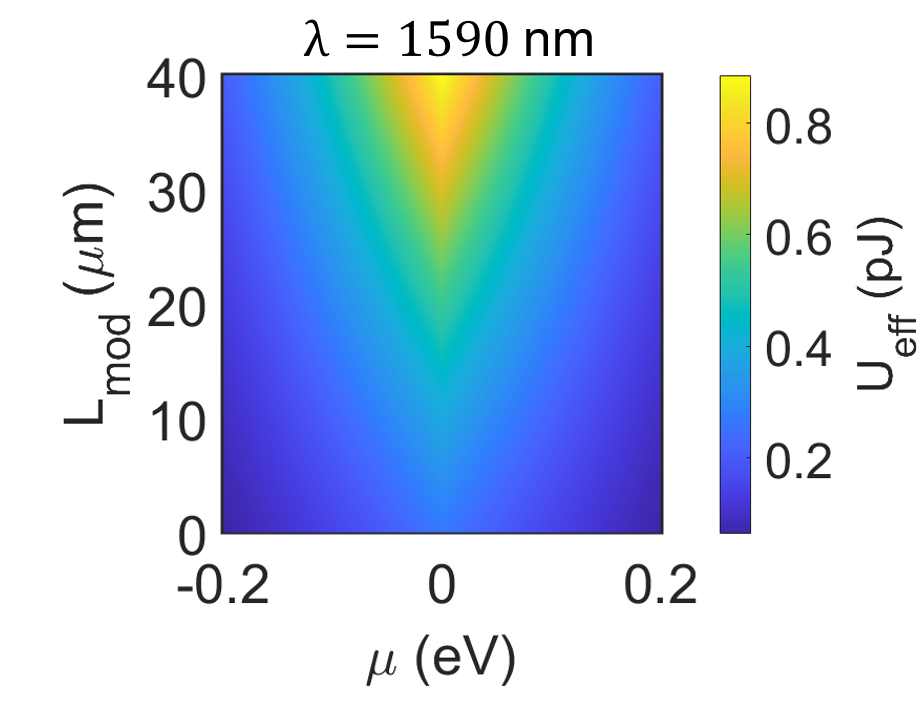}
  \caption{}
  \label{fig:1590}
\end{subfigure}
\caption{(a) Effective switching energy ($U_{\text{eff}}$) as a function of chemical potential ($\mu$) and modulator length ($L_{\text{mod}}$). $\lambda_{\text{pump}}=$ 1570$\,$nm. (b) $U_{\text{eff}}$ as a function of chemical potential and modulator length. $\lambda_{\text{pump}}=$ 1590$\,$nm.}
\label{fig:tT00O}
\end{figure}

\begin{figure}
  \centering
  \includegraphics[width=0.45\linewidth]{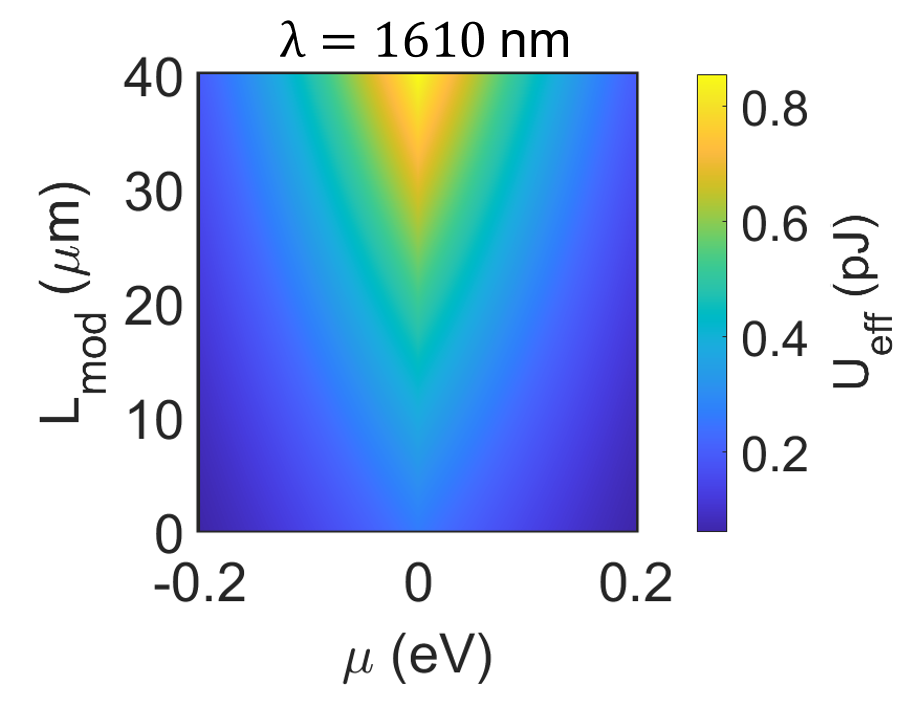}
\caption{Effective switching energy ($U_{\text{eff}}$) as a function of chemical potential ($\mu$) and modulator length ($L_{\text{mod}}$). $\lambda_{\text{pump}}=$ 1610$\,$nm.}
\label{fig:1610}
\end{figure}

The effective switching energy ($U_{\text{eff}}$) as a function of the modulator length ($L_\text{mod}$) and chemical potential ($\mu$) at the pump wavelengths 1490, 1510, 1530, 1550, 1570, 1590 and 1610$\,$nm is presented in Figs. \ref{fig:totITO}, \ref{fig:totT1O}, \ref{fig:tT00O} and \ref{fig:1610}.

\newpage

\section{Modeling Parameters}

The refractive index data of hBN are taken from ref. \cite{rah2019optical}. Graphene is modeled in Lumerical as a 2D material with a surface optical conductivity ($\Tilde{\sigma}$) that is given by \cite{grapheneModel, hanson2008dyadic}:

\begin{equation} \label{eq:1st}
    \Tilde{\sigma}(\omega, \Gamma, \text{\textmu}, T) = \Tilde{\sigma}_{intra}(\omega, \Gamma, \text{\textmu}, T) + \Tilde{\sigma}_{inter}(\omega, \Gamma, \text{\textmu}, T)
\end{equation}

\begin{equation} \label{eq:2nd}
     \Tilde{\sigma}_{intra}(\omega, \Gamma, \text{\textmu}, T) = \dfrac{-je^2}{\pi\hbar^2(\omega+j2\Gamma)}\int_{0}^{\infty}E\,(\dfrac{\partial f(E)}{\partial E} - \dfrac{\partial f(-E)}{\partial E}) \: dE
\end{equation}

\begin{equation} \label{eq:3rd}
     \Tilde{\sigma}_{inter}(\omega, \Gamma, \text{\textmu}, T) = \dfrac{je^2(\omega+j2\Gamma)}{\pi\hbar^2}\int_{0}^{\infty}\dfrac{f(-E) - f(E)}{(\omega+j2\Gamma)^2 - 4(E/\hbar)^2} \: dE
\end{equation}

\begin{equation} \label{eq:4th}
     f(E) = (e^{(E-\mu)/k_{B}T}+1)^{-1}
\end{equation}

\noindent where $\Tilde{\sigma}_{intra}$ and $\Tilde{\sigma}_{inter}$ account for the surface optical conductivity due to intraband and interband absorption, respectively. $\omega$ is the angular frequency of incident photons, $\Gamma$ is the scattering rate of graphene, $T$ is the operation temperature, $e$ is the electron charge, $\hbar$ is the reduced Planck constant, $f(E)$ is the Fermi-Dirac distribution, and $k_{B}$ is the Boltzmann constant.

\bibliographystyle{IEEEtran}
\bibliography{references}